\documentclass[a4paper]{jpconf}
\usepackage{graphicx}
\begin{document}

\title{\bf Vibron Self--trapped States in Biological Macromolecules:
Comparison of Different Theoretical Approaches}

\author{ D. \v Cevizovi\' c$^{1,2}$, S. Galovi\' c$^{1,2}$, \underline{A.
Reshetnyak}$^3$, and Z. Ivi\' c$^1$}

\address{$^1$ University of Belgrade, "Vin\v ca" Institute of Nuclear Sciences,
Laboratory for theoretical and condensed matter physics, Belgrade, Serbia\\
$^2$ Joint Institute for Nuclear Research, Bogoliubov Laboratory for
Theoretical Physics, Dubna, Russia\\
$^3$ Laboratory of Non--Linear Media Physics, Institute of Strength Physics and
Material Science SB RAS,  Tomsk 634021, Russia\\}

\ead{reshet@ispms.tsc.ru}

\begin{abstract}
 A study of the applicability of the variational treatments based on using of the modified Lang-Firsov unitary transformation (MLF method) in the investigation of the vibron self-trapped states in biological macromolecular chains are presented. We compare  the values of the ground state energy
predicted by MLF methods with the values of the ground state energy predicted by
the standard small--polaron theory, for various values of the basic energy
parameters of the system. We obtain regions in system parameter space where MLF
approach gives better description of the vibron states.
\end{abstract}

\section{Introduction}

There exist  many unresolved fundamental problems in molecular biology and biochemistry  and one from them is the problem of long--distance energy transport in biological macromolecules.
It is believed that the energy released by the hydrolysis of adenosine
triphosphate (ATP) appears by an universal energy source allowing many biological
processes. Mentioned energy released by the ATP hydrolysis may be captured by the
protein molecules and excite the high--frequency amide--I vibration of a
peptide group. Due to the dipole--dipole coupling between the neighbouring
peptide groups the excited amide--I vibration is delocalized causing formation
of the vibron state which, in turn, can move along the polypeptide chain. However, it is yet
not clearly understood how this energy can be transported along macromolecule at
long distances, without being dissipated or dispersed. An earlier explanation of
this process is based on the soliton theory \cite{Davydov}, \cite{Davydov1}, \cite{Davydov2}. According to those
models, a soliton arises due to the self--trapping (ST) of the amide--I quanta.

According to the general theory of ST phenomena \cite{Rashba}, ST problem
exhibits two asymptotic solutions depending on the values of three basic energy
parameters of the system: the vibron bandwidth $2\left|J\right|$, the phonon
characteristic frequency $\omega_C$ and the SP bending energy $E_b$ (which is
proportional to the strength of the vibron--phonon coupling). Thus, in the
adiabatic limit ($2\left|J\right|>\hbar\omega_C$) lattice distortion with large
inertia can not  follow the exciton motion and it forms an essentially static
potential well where that particle may be trapped. Depending on the value of the
mutual ratio of $E_b$ and $\hbar\omega_C$, the spatial extent of an entity
created in this way (polaron) may vary from being concentrated around one site
only ($E_b>>\hbar\omega_C$) i.e. adiabatic SP (ASP), while in the case when
$E_b<\hbar\omega_C$ it is spread over the large number of lattice sites and we
have adiabatic large polaron (ALP) or soliton. By contrast, in non--adiabatic
limit ($2\left|J\right|<\hbar\omega_C$) the quantum nature of the phonons plays
a crucial role. In such case a vibron is dressed by a virtual cloud of phonons
which yields a lattice distortion essentially located on a single site and
instantaneously follows the vibron motion. The vibron dressed by the virtual
phonon cloud forms a small polaron (SP) whose properties are well described by
performing the so--called Lang--Firsov (LF) unitary transformation \cite{LF}. As
a consequence, conventional SP theories are applicable in the strong--coupling,
non--adiabatic limit.

Unfortunately, the values of the basic energy parameters in biological
macromolecular chains belong to non--adiabatic limit. For that reason, it has
been suggested by some authors \cite{AKPRB33},\cite{BIPRB40}, \cite{BIPRB48} that vibron
self--trapping in hydrogen--bonded macromolecular chains might result in the
formation of a small--polaron, rather than a soliton. From the other side, the
values of the energy parameters indicate that the strength of vibron--phonon
coupling in hydrogen--bonded macromolecules falls in the weak to intermediate
limits \cite{PouthierJCP132}. In addition, some recent numerical \cite{HammPRB78}
studies of the ST phenomena in hydrogen--bonded macromolecular chains indicate
that its proper theoretical description requires an approach that goes beyond
the conventional strong--coupling SP theories. Namely, according to results from
\cite{HammPRB78} it seems that depending on the temperature and the values of
the system parameters the abrupt transition from partially dressed (light and
mobile) to self--trapped (practically immobile quasi particle) may occur. This
situation can not be described in the framework of the standard SP approach
(based on  using of the Lang--Firsov unitary transformation) and as a
consequence its theoretical investigations require the means which
involve the concept of partial dressing. This concept is based on modified
Lang--Firsov (MLF) unitary transformation and  is applicable in a wide
part of system parameter space. Such an approach is considered in a close correspondence
with the supplementary variational treatment of the problem by means of the
Toyozawa ansatz \cite{ToyozawaPTP26}. In our recent paper
\cite{CevizovicPRE84} we used slightly more flexible method in order to
investigate the temperature dependence of the SP states character.

In this paper we consider two variational formalisms based on  using of the
modified Lang--Firsov approach and compare obtained results with ones
predicted by the standard SP theory. The  variational parameter(s) introduced here
characterizes the degree in which the vibron distorts the lattice and the lattice
feedback on the vibron, i.e. vibron dressing. The first variational approach
(so called ``$f_q$--approach'') is based on the model used in \cite{CevizovicPRE84}, while the
second one (``$\delta$--approach'') exploits the method introduced by Brown
and Ivi\'c in \cite{BIPRB40}, \cite{BIPRB48} in the form used in \cite{IvicPD113}. The last approach
intermediates between non-adiabatic and adiabatic limit with use of only one
variational parameter. We calculate and compare the system ground state energies
for various basic system parameter values in the case when vibron interacts with
both optical and acoustical phonon subsystems. Our attention was restricted to
the single--vibron case.

\section{Theoretical methods}

The system under consideration consists of single vibron excited on $n$--th
structural element of the macromolecule which, in turn, interacts with thermal
oscillations of the chain through the linear short--ranged deformation
potential interaction. We suppose that vibron excitation can move along the
chain from $n$--th to its nearest--neighbouring structural element. The corresponding
Hamiltonian can be written in the following form \cite{HolsteinAP8},
\begin{equation}\label{PocHam}
H=\Delta\sum_n{a^{\dagger}_na_n}-\sum_n{Ja^{\dagger}_n
(a_{n+1}+a_{n-1})}+\sum_q{ \hbar\omega_qb^{\dagger}_qb_q}+\frac{1}{\sqrt{N}}
\sum_q\sum_n{F_q\mathrm{e}^{iqnR_0}a^{\dagger}_na_n(b_q+b^{\dagger}_{-q})},
\end{equation}
where $\Delta$ is the vibron excitation energy, $a^{\dagger}_n(a_n)$ describes
the presence (absence) of the vibron quanta on the structural element, which is
positioned on $n$--th lattice site, $b^{\dagger}_q(b_q)$ creates (annihilates)
phonon quanta, and $\omega_q$ is the phonon frequency. The inter--site overlap
integral $J$ characterize the vibron transfer between neighbouring structural
elements in the chain. Finally, $F_q=F^*_{-q}$ is the vibron--phonon coupling
parameter which governs the character of ST states. In order to investigate the
vibron self--trapping in macromolecular chains, we shall consider the following
quasi particle--phonon interaction:
\begin{enumerate}
  \item[(a)] interaction with dispersionless optical phonon modes
\cite{HolsteinAP8},\cite{DavydovTTT}, \cite{Mahan}
$$F_q=\chi\sqrt{\frac{\hbar}{2M\omega_q}},$$
where $\chi$ is the vibron--phonon coupling constant, which accounts the
influence of the external motion of the vibron on the $(n\pm 1)$--th structural
elements and  on the modulation of the vibron frequency in the $n$--th structural
element, $\omega_q=\omega_C$ is phonon dispersion law,
$\omega_C=2\sqrt{\kappa/M}$, $\kappa$ appears by the stiffness of the chain, $M$ is the
molecular group mass;
  \item[(b)] interaction with acoustic phonon modes through the short--ranged deformation
potential (acoustic deformation potential, ADP)
\cite{DavydovTTT}, \cite{Mahan}
$$F_q=2i\chi\sqrt{\frac{\hbar}{2M\omega_q}}\sin{qR_0},$$
with $\omega_q=\omega_C\sin{\left|qR_0/2\right|}$ being by phonon dispersion law, and
$R_0$ is a lattice constant.
\end{enumerate}

In order to examine the degree of the vibron dressing, we  pass to the
small--polaron picture using unitary transformation operator, $$U=\exp\left\{
-\frac{1}{\sqrt{N}}\sum_q\sum_n{f_q\mathrm{e}^{-iqnR_0}a^{\dagger}_na_n(b_{-q}
-b^{\dagger}_q)}\right\}.$$
Here $f_q=f^*_{-q}$ are variational parameters,
while operators $a^{\dagger}_n(a_n)$ represent (partially) dressed vibrons. In
the case of the passing to the standard small--polaron picture, parameter $f_q$
has standard Lang--Firsov form: $f_q=\frac{F^*_q}{\hbar\omega_q}$. Operators
$a^{\dagger}_n(a_n)$ that appear in the transformed Hamiltonian, represent the
(fully) dressed quasi particle. In all cases the operators $b^{\dagger}_q(b_q)$
describe the new phonons in the lattice with shifted equilibrium positions
of molecular groups.

Partially dressed vibron states represent the dynamically stable  eigen-states of
the system provided that these variational parameters correspond to the
minimum energy state. Thus, we search for their optimized values minimizing the
system ground--state energy. According to the method described in
\cite{YarkonySilbeyJCP65}, optimization procedure will be performed in a
mean--field manner, which gives better upper bound of the system free energy
and consequently the better upper bound of the system ground state energy. In
particular, we define (in wave number representation) an effective, mean--field
Hamiltonian $\tilde{H}_0$ in the following way,
\begin{equation}\label{MFH}
\tilde{H}=\tilde{H}_0+\tilde{H}_{int},
\end{equation}
where $\tilde{H}_0=\tilde{H}_v+\tilde{H}_{ph}$, $\tilde{H}_v=\left\langle
\tilde{H}-\tilde{H}_{ph}\right\rangle_{ph}$ and $\tilde{H}_{int}=\tilde{H}
-\tilde{H}_{ph}-\left\langle\tilde{H}-\tilde{H}_{ph}\right\rangle_{ph}$. The
symbol $\left\langle \ \right\rangle_{ph}$ denotes the average over the new--phonon
ensemble, which is considered in the thermal equilibrium state at the temperature $T$. In such a
way we derive,
\begin{equation}
\tilde{H}_0=\sum_k{E_{SP}(k)a^{\dagger}_ka_k}+\sum_q{\hbar\omega_q
b^{\dagger}_qb_q},
\end{equation}
where $a_k=\frac{1}{\sqrt{N}}\sum_n{\mathrm{e}^{iknR_0}a_n}$, and
\begin{equation}\label{ESP}
E_{SP}(k)=\Delta-\frac{1}{N}\sum_q{\left\{(f_q+f^*_{-q}
)F_q-\hbar\omega_q\left|f_q\right|^2\right\}}-2J\mathrm{e}^{-W(T)}\cos(kR_0)
\end{equation}
denotes the variational energy of the small--polaron band states, while
$W(T)=\frac{1}{N}\sum_q{\left|f_q\right|^2(2\bar{\nu}_q+1)(1-\cos(qR_0))}$
denotes the vibron--band narrowing factor, which characterizes the degree of
the reduction of the corresponding overlap integral or equivalently the
enhancement of the polaron effective mass. For obvious reasons, they are
sometimes called the "`dressing fractions'" or the "`dressing parameters'".
At last, the quantity $\bar{\nu}_q$, $\bar{\nu}_q=1/(\mathrm{e}^{\hbar\omega_q/k_BT}-1)$, denotes the phonon
average number.

The polaron--phonon interaction term may be neglected provided that dressed vibrons
represent sufficiently well eigen states of system. In such a way it cannot affect SP
equilibrium properties substantially and at $T=0$ K  the further procedure simply
reduces to minimization of the SP ground state energy. We note the
ground--state vector of the effective Hamiltonian $\tilde{H}_0$ reads, $\left|
\psi_{GS}\right\rangle=a^{\dagger}_k\left|0\right\rangle_{vib}\otimes\prod_q
\left|0\right\rangle_q$, where $\prod_q\left|0\right\rangle_q$ is the phonon
vacuum vector so that the ground--state energy corresponds to the lowest level
of SP energy, i.e.,$E_{GS}=\left\langle\psi_{GS}\left|\tilde{H}_0\right|
\psi_{GS}\right\rangle =E_{SP}(k)$. Thus, from the  requirement, $\left.\frac{\partial
E_{SP}(k)}{\partial f_q}\right|_{k_0}=0$, we get,
\begin{equation}\label{fq}
f_q=\frac{F^*_q}{\hbar\omega_q+4|J|\mathrm{e}^{-W}\left(2\bar{\nu}_q+1\right)
\sin^2\left(\frac{qR_0}{2}\right)}.
\end{equation}
From the Eq.(\ref{fq}) it follows that for each value of $q$ there is the corresponding value of the
variational parameter $f_q$. For that reason, the optimization procedure for
this approach becomes extremely difficult problem, even numerically. Note, however,
that only $k_0=\pi/R_0$ (when $J<0$), or $k_0=0$ (when $J>0$), should be
considered since it will correspond to the polaron ground state only. This case is also the most
relevant for the optical spectroscopy since vibron bandwidths in these media are
very narrow and its dispersion in all practical examination of optical spectra
can be neglected \cite{HammPRB78}, \cite{DavydovTTT}, \cite{Mahan} \cite{EminAP22}, \cite{ScottPR217}.

The expressions for vibron ground state energy (normalized on the characteristic
phonon energy $\hbar\omega_C$ and measured from the energy level $\Delta$) and
vibron dressing factor are obtained by the substitution of the variational
parameters into (\ref{ESP}). Performing the summation over the phonon quasi
momenta $q$ by virtue of the rule $\frac{1}{N}\sum_q{\dots}\rightarrow
\frac{R_0}{2\pi}\int^{\pi/R_0}_{-\pi/R_0}{\dots dq}$ and by introducing two
dimensionless parameters (that determine the system parameter space)
$S=\frac{E_b}{\hbar\omega_C}$ and $B=\frac{2\left|J\right|}{\hbar\omega_C}$, we
obtain,
\begin{equation}
\mathcal{E}_{GS}=-W-3BW\mathrm{e}^{-W}-B\mathrm{e}^{-W},
\end{equation}
and
\begin{equation}
W=\frac{S}{(1+2B\mathrm{e}^{-W})^{3/2}}
\end{equation}
in the case when vibron interact with optical phonon mode, and
\begin{equation}
\mathcal{E}_{GS}=-\frac{W}{2}\frac{I_E}{I_W}-B\mathrm{e}^{-W},
\qquad
W=8SI_W,
\end{equation}
when the vibron interacts with acoustic phonon modes. The quantities $I_E, I_W$  that appear in the
above relations have the integral representation,
$$I_E=\frac{1}{\pi}\int^{\pi/2}_0{\cos^2x\frac{1+4B
\mathrm{e}^{-W}\sin x}{(1+2B\mathrm{e}^{-W}\sin x)^2}dx}\texttt{, and }I_W=\frac{1}
{\pi}\int^{\pi/2}_0{\frac{\sin x\cos^2x}{(1+2B\mathrm{e}^{-W}\sin x)^2}dx}.$$

In the case of simple ``$\delta$--approach'' we choose $f_q$ in the form:
$f_q=\frac{\delta F^*_q}{\hbar\omega_q}$, where $0<\delta<1$ is variational
parameter. In that case, we have more simple variational technique in which  we should determine only
one variational parameter ($\delta$). On the other hand, this
assumption implies the equal dressing for all phonon modes. At the first sight, it
looks like a very strong assumption since the whole set of variational
parameters (one for each mode) is substituted by a single one. However, here we
showed that predictions obtained by this approach are very close with ones,
obtained by more rigorous ``$f_q$ approach''. Repeating above described
procedure we obtained condition which determines parameter $\delta$,
\begin{equation}\label{delta}
4J\delta W_{LF}\mathrm{e}^{-\delta^2W_{LF}}-2E_b(1-\delta)=0,
\end{equation}
where $E_b=\frac{1}{N}\sum_q{\frac{\left|F_q\right|^2}{\hbar\omega_q}}$ is
small--polaron binding energy, while $W_{LF}=\frac{1}{N}\sum_q
{\frac{\left|F_q\right|^2}{(\hbar\omega_q)^2}(1-\cos qR_0)}$ is SP band
narrowing factor calculated in the non--adiabatic, strong coupling limit
(standard LF approach). Then, performing summation over phonon quasimomenta,
we derive the relation for the ground state energy (in $(S,B)$ parameter space) in the form,
\begin{equation}
\mathcal{E}_{GS}=-\delta(2-\delta)S-B\mathrm{e}^{-\delta^2S}
\end{equation}
in the case when vibron interact with optical phonon, and
\begin{equation}
\mathcal{E}_{GS}=-\delta(2-\delta)S-B\mathrm{e}^{-\frac{8\delta^2S}{3\pi}},
\end{equation}
when vibron interact with acoustical phonons.

\section{Results}

According to the available literature data, the hopping constant in
$\alpha$--helix proteins is typically equal to, $J=7.8\;\mathrm{cm}^{-1}$ (in the case of
the hydrogen bonds between peptide units) and $J=-12.4\;\mathrm{cm}^{-1}$
(between different spines of hydrogen--bonded peptide units, i.e. in
the case of the covalent bonds between peptide unit)
\cite{ScottPR217,ScottPRA26}. The mass of the peptide unit ranges between
$M=2\cdot10^{-25}$ kg and $M=5.7\cdot10^{-25}$ kg
\cite{PouthierJCP132,ScottPR217}.

The above mentioned values of the system parameters in macromolecular chains
indicate that adiabatic parameter in these structures ranges between $B=0.01$
and $B=0.5$. In the same time, coupling constant ranges between $S=0.01$ and
$S=0.3$. The last fact indicates that these structures belong to the non-adiabatic coupling limit from weak to
intermediate one.

Obtained numerical results are presented on the Figs. 1--4. The calculations
obtained by ``$\delta$--approach'' are presented by full lines, the ones
obtained by the ``$f_q$--approach'' are presented by dashed lines, and the
calculations obtained by the standard LF approach are presented by dotted lines.
As one can remark, the results obtained by both variational approaches are
strongly different compared with ones obtained by standard LF approach, for
small values of coupling constant $S$. This difference disappears on large
values of $S$ for all values of $B$. For small and intermediate values of $S$,
the difference between two variational approaches increases with increasing of
$B$. These results are expectable, since standard LF approach is not applicable
in weak coupling limit. On the other side, results obtained by both variational
approaches are very similar in the whole range of $S$, for small values of $B$
(non-adiabatic limit). There are significant difference only for $B>1$ (adiabatic
limit), for small and intermediate values of $S$.

One should be noted in addition,  it is evident that both variational approaches predict a
multivalued dependence of $\mathcal{E}_{GS}$ and $W$ on $S$, in certain region
of $S$, only in the adiabatic limit. According to the variational principle, the only
lowest energy value is physically meaningful. Consequently, only lower
values of $\mathcal{E}_{GS}$ (and its corresponding values of $W$) have
physical significance and truly variational energy is presented by one line
which is not smooth (at certain value of $S$ it have the point of cusp).
In the same time, $W$ undergoes an abrupt change. That means that vibron ST
state have abrupt transition from weakly dressed to heavy dressed polaron
state. Obtained results may look as a consequence of the applied variational
approach. However, it seems that these results are in good agreement with some
recent numerical investigations \cite{HammPRB78}, where authors interpreted
similar results as a coexistence of two types of SP states instead of abrupt
transition from one to another quasiparticle type.

\begin{figure}[h]
\begin{minipage}{28pc}
\includegraphics[width=28pc]{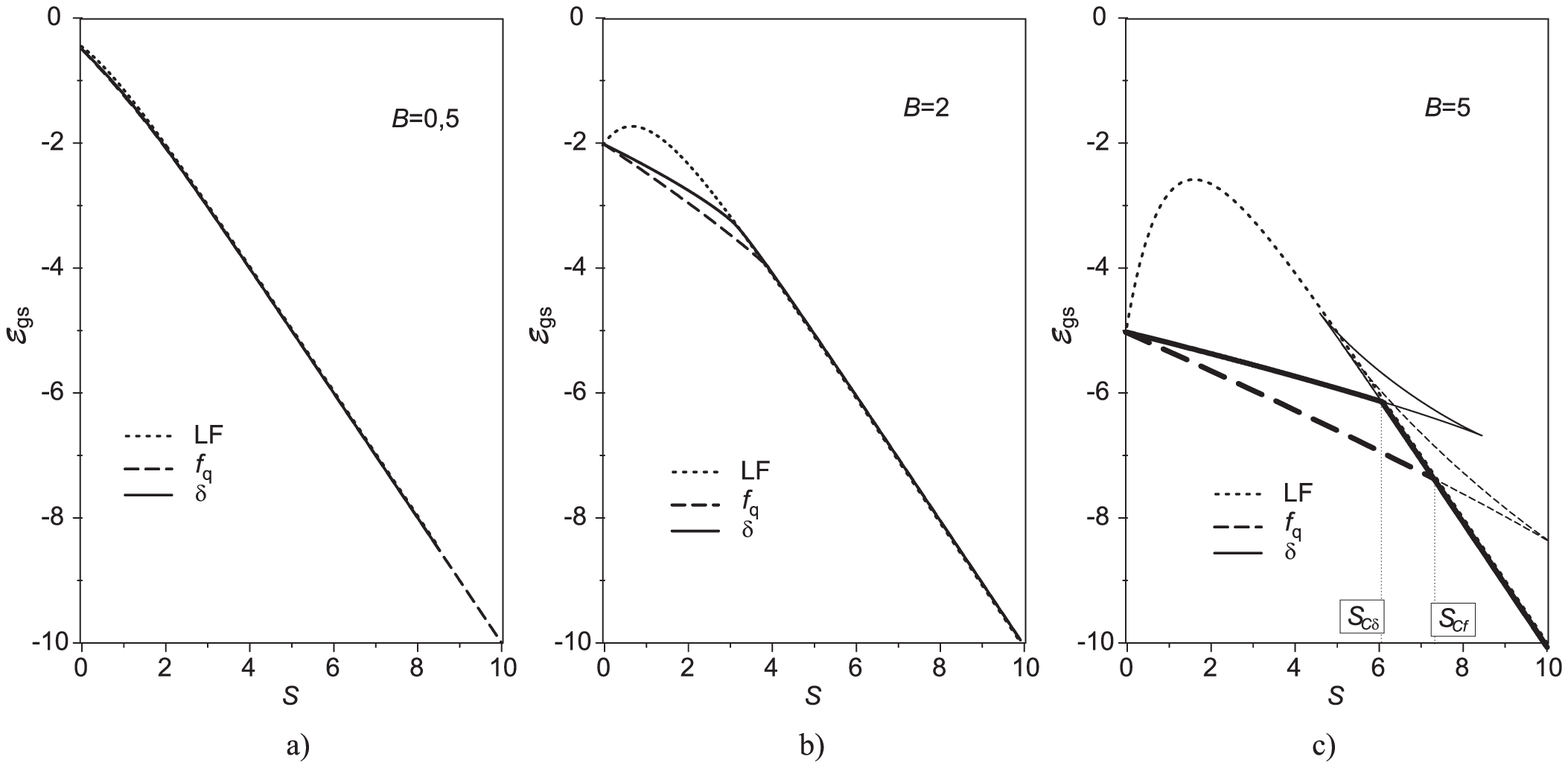}
\caption{\label{Fig1a} The dependence of the $\mathcal{E}_{GS}$ versus $S$,
for different values of the $B$. The case of vibron that interacting with
optical phonon modes.}
\end{minipage}\hspace{2pc}\\
\begin{minipage}{28pc}
\includegraphics[width=28pc]{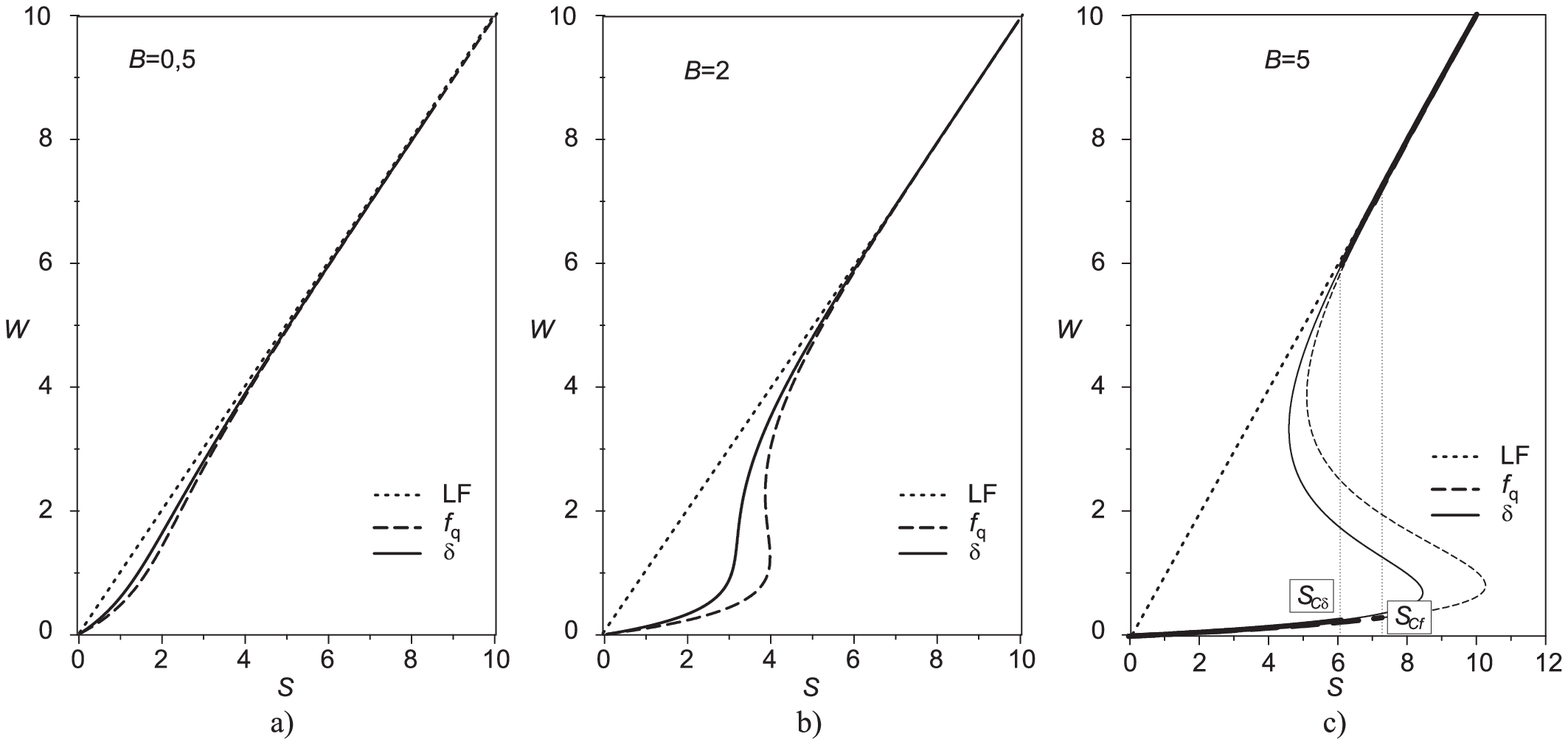}
\caption{\label{Fig1b} The dependence of the $W$ on the $S$, for
different values of the $B$. The case of vibron that interaction with
optical phonon modes.}
\end{minipage}
\end{figure}

\begin{figure}[h]
\begin{minipage}{28pc}
\includegraphics[width=28pc]{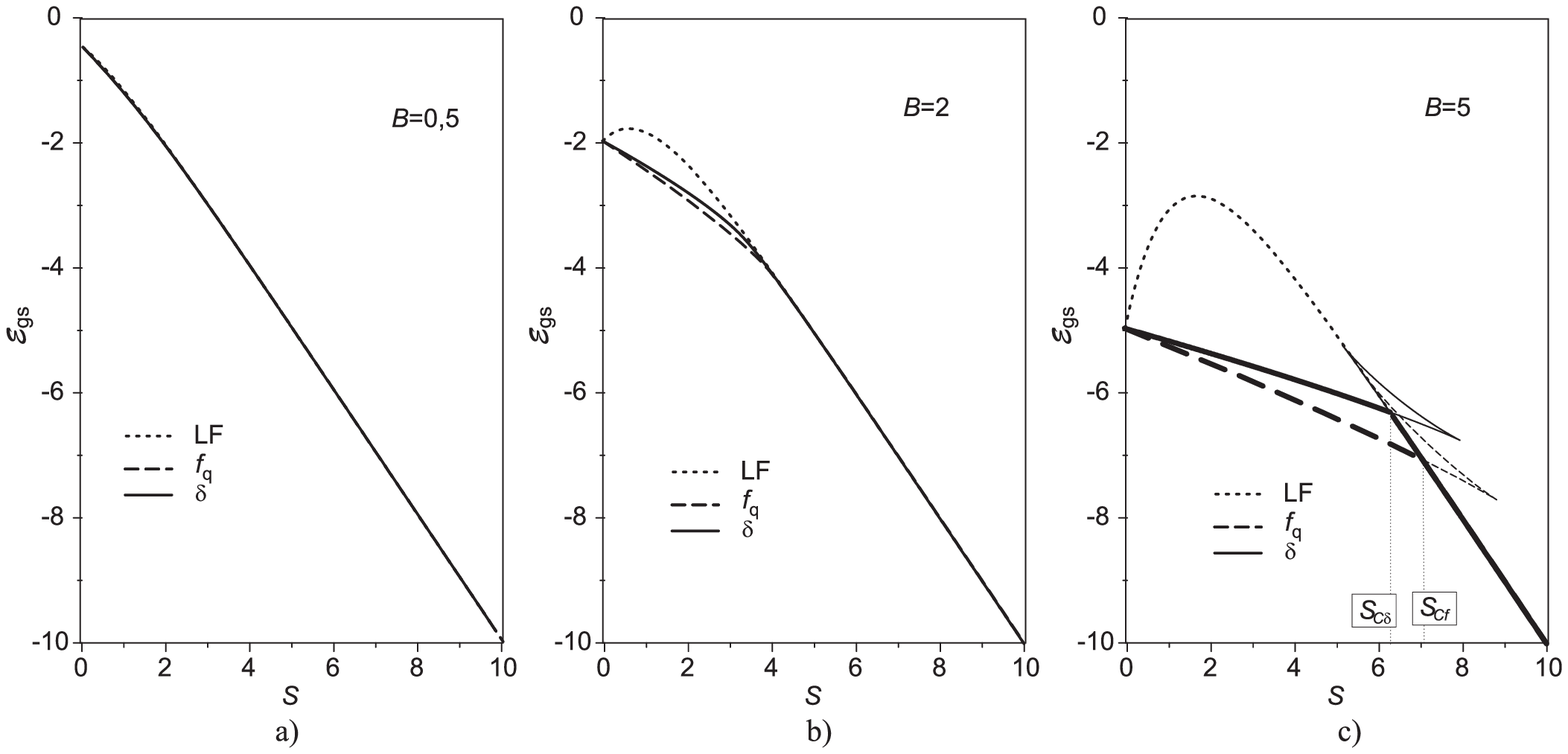}
\caption{\label{Fig2a} The dependence of the $\mathcal{E}_{GS}$ versus $S$,
for different values of the $B$. The case of vibron that interacting with
acoustic phonon modes.}
\end{minipage}\hspace{2pc}\\
\begin{minipage}{28pc}
\includegraphics[width=28pc]{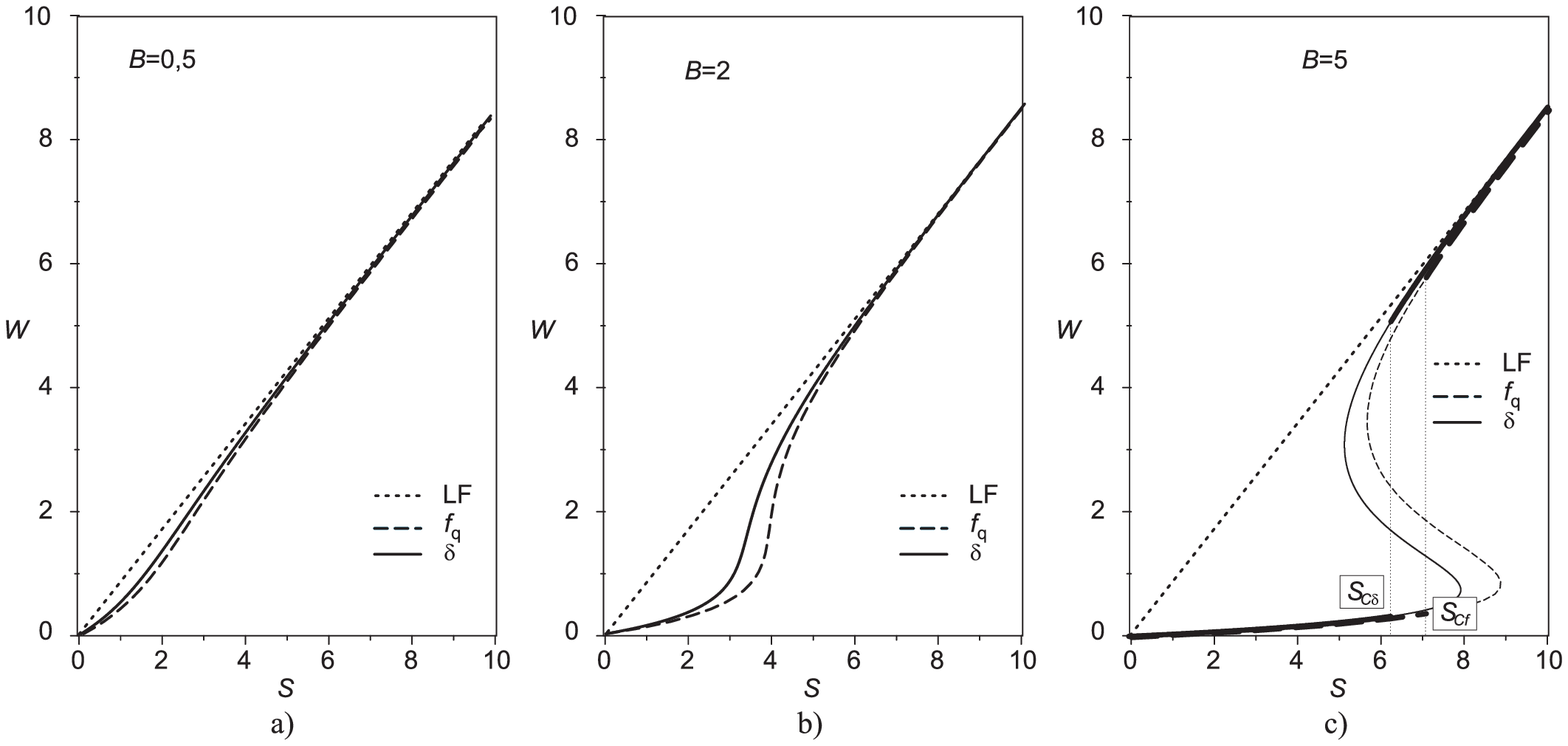}
\caption{\label{Fig2b} The dependence of the $W$ on the $S$, for
different values of the $B$. The case of vibron that interaction with
acoustic phonon modes.}
\end{minipage}
\end{figure}

\section{Conclusions}

In accordance  to the obtained results, one may note that both variational approaches
give very similar qualitative and quantitative predictions, especially in
non-adiabatic region. With the rise of the adiabatic parameter, the difference
between variational approaches, as well as between variational  approaches and
standard small--polaron approach vanishes. In the case of the vibron interacting
with the optical phonon modes, ground state energies predicted by the
``$f_q$--approach'' are something lower than ground state energies predicted by
the ``$\delta$--approach''. But, in the non--adiabatic region of the parameter
space this difference may be practically neglected, and ``$\delta$--approach''
is quite good method for the investigation of the vibron self--trapping states.
Similar results are obtained in the case when vibron interacts with acoustic
phonon modes. In the adiabatic region of the system parameter space the difference
between two variational approaches is significant.

Due to the fact that practically all biological macromolecular structures
belongs to the non-adiabatic, weak or intermediate coupling limit, it seems that
simple ``$\delta$--approach'' may be by the hopeful method for investigations of the
self--trapped vibron states in such structures.

\subsection*{Acknowledgments}

The work was supported by the Ministry of Education and Science of Republic of
Serbia under contract numbers III--45005 and III--45010, and by the project
within the Cooperation Agreement between the JINR, Dubna, Russian Federation and
Ministry of Education and Science of Republic of Serbia.


\section*{References}
\begin{thebibliography}{9}
\bibitem{Davydov}
 Davydov A S 1991 {\it Solitons in Molecular Systems} (Dordrecht: Kluwer Academic Publishers)

\bibitem{Davydov1}
 Davydov A S 1981 The role of solitons in the energy and electron transfer in
one -dimensional molecular systems \textit{Physica} D {\bf 3} 1

\bibitem{Davydov2}
Davydov A S 1982 {\it Biology and Quantum Mechanics} (New York: Pergamon Press)

\bibitem{Rashba}
 Rashba E I 1982  {\it Excitons}  ed E I Rashba and  M D Struge  (Amsterdam: North Holland) p 543

\bibitem{LF}
 Lang I J and Firsov Yu A  1962  Sov. Phys. JETP \textbf{16}  1301

\bibitem{AKPRB33}
 Alexander D M and  Krumhansl J A 1986 Localized excitations
in hydrogen--bonded molecular crystals \textit{Phys. Rev.} B {\bf 33} 7172

\bibitem{BIPRB40}
 Brown D W and  Ivi\'c Z 1989   Unification of polaron and soliton
theories of exciton transport \textit{Phys. Rev.} B {\bf 40} 9876

\bibitem{BIPRB48}  Ivi\'c Z,  Kapor D,
Skrinjar M and Popovi\'c Z 1993 Self--trapping in
quasi--one--dimensional electron-- and exciton--phonon systems \textit{Phys. Rev.} B {\bf 48} 3721

\bibitem{PouthierJCP132}
 Pouthier V  2010 Vibron phonon in a lattice of H-bonded
peptide units: A criterion to discriminate between the weak and the strong coupling limit \textit{J. Chem. Phys.}   {\bf 132} 035106

\bibitem{HammPRB78}
 Hamm P and  Tsironis G 2008 Barrier crossing to the small
Holstein polaron regime \textit{Phys. Rev.} B  {\bf 78} 092301

\bibitem{ToyozawaPTP26}
 Toyozawa Y 1961 Self--Trapping of an Electron by the
Acoustical Mode of Lattice Vibration I \textit{Progr. Theor. Phys.}  {\bf 26} 29

\bibitem{CevizovicPRE84}
 \v Cevizovi\' c D,  Galovi\' c S and  Ivi\' c Z 2011 Nature
of the vibron self--trapped states in hydrogen--bonded macromolecular chains \textit{Phys. Rev.} E
{\bf 84} 011920

\bibitem{IvicPD113}
 Ivi\'c Z 1998 The role of solitons in charge and energy transfer in
$1D$ molecular chains \textit{Phys.} D {\bf 113} 218

\bibitem{HolsteinAP8}
 Holstein T 1959 Studies of Polaron Motion \textit{Ann. Phys.}  {\bf 8} 325


\bibitem{DavydovTTT}
 Davydov A S 1976 {\textit{Theory of solids}} \textit{(Teoriya tverdogo tela)}, (Moscow: Nauka, on Russian)

 \bibitem{Mahan} Mahan G D 1990 {\textit{Many--Particle Physics}} (New York:Plenum Press)

\bibitem{YarkonySilbeyJCP65}  Yarkony D and  Silbey R 1976 Comments on exciton phonon coupling: Temperature dependence  \textit{J. of
Chem. Phys.}
{\bf 65} 1042

\bibitem{EminAP22}
 Emin D 1973 On the existence of free and self--trapped
carriers in insulators: an abrupt temperature--dependent conductivity
transition  \textit{Advan. In. Phys.}  {\bf 22} 57


\bibitem{ScottPR217}  Scott A C 1992 Davydov's soliton \textit{Phys. Rep.}   {\bf 217} 1

\bibitem{ScottPRA26} Scott A C 1982  Dynamics of Davydov
solitons \textit{Phys. Rev.} A  {\bf 26} 578


\end{thebibliography}
\end{document}